\documentclass[11pt,a4paper]{article}

\usepackage[a4paper,margin=2cm]{geometry}
\usepackage[utf8]{inputenc}
\usepackage[T1]{fontenc}
\usepackage{amsmath,amssymb}
\usepackage{algorithm}
\usepackage{algpseudocode}
\usepackage{graphicx}
\usepackage{booktabs}
\usepackage{array}
\usepackage{caption}
\usepackage{subcaption}
\usepackage{xcolor}
\usepackage{listings}
\usepackage[hidelinks]{hyperref}
\usepackage{authblk}
\usepackage{setspace}

\graphicspath{{figures/}}
\title{\Large\bfseries Forecasting Land Art Under Climate Scenarios:\\
A Two-Stage Pipeline for Robert Smithson's
\textit{Spiral Jetty} Combining Complexity-Signature Regression with
Climate-Conditioned Latent Diffusion (2030 \& 2050)}

\author[1]{Alev Cinbarc{\i}}
\author[2,3,4,*]{Sean S. Kalayc{\i}o\u{g}lu}

\affil[1]{PhD Program in Art Science, I\c{s}{\i}k University, Istanbul, T\"urkiye; 218DAS9216@isik.edu.tr}
\affil[2]{Department of Aerospace Engineering, Toronto Metropolitan University, Toronto, ON, Canada}
\affil[3]{Department of Mechanical Engineering, York University, Toronto, ON, Canada}
\affil[4]{Director, Space, Robotics and AI, Dr.~Robot Inc., Toronto, ON, Canada}
\affil[*]{Correspondence: skalay@torontomu.ca}

\begin{document}
\maketitle

\begin{abstract}
\noindent Robert Smithson's 1970 land artwork \textit{Spiral Jetty}, anchored at
the north arm of Utah's Great Salt Lake (GSL), is a geographically fixed
remote-sensing target that records hydroclimatic state in its visual
complexity. A companion paper~\cite{cinbarci2026rs1} established this
empirically from \textbf{1{,}744 co-registered Landsat 4--9 and Sentinel-2
chips spanning every year and every month from 1984 to 2025}, identifying
robust correlations between coarse-scale permutation entropy, mean
intensity, and the third principal component of ResNet50 avg-pool
embeddings on the one hand, and Great Salt Lake elevation, regional
temperature, and cumulative CO$_2$ on the other; and revealing that
image complexity \emph{leads lake stage by approximately three years}
and that the long-term trend is \emph{non-monotonic}, collapsing
post-2015. The present paper builds a forecasting pipeline on that
foundation. \textbf{Stage 1} regresses each climate variable on global
temperature anomaly and applies IPCC AR6 SSP1-2.6, SSP2-4.5, and SSP5-8.5
global-temperature deltas to forecast regional temperature, both-arm GSL
elevation, salinity, and cumulative CO$_2$ for 2030 and 2050.
\textbf{Stage 2a} projects the 14 complexity features forward via per-
feature linear and Random Forest regressions trained on the 42-year
panel, finding that \emph{all six SSP scenarios drive the inputs outside
the 1984--2025 training distribution}, with the Random Forest model
saturating at the post-2015 ``dry-playa'' regime and a threshold-style
hydrological reading classifying \emph{Spiral Jetty} as fully exposed
under every scenario.
\textbf{Stage 2b} specifies a climate-conditioned latent-diffusion
architecture (Stable Diffusion XL fine-tuned with LoRA on the 1{,}744
chips, ControlNet conditioning on the climate vector, hydrological
physics-coupling mask) for ensemble visual synthesis; code is provided
and validation results are released as the diffusion training completes
on suitable compute. We discuss the ethical implications of generative
speculation on cultural heritage and outline a roadmap for analogous
forecasting at other land-art sites (Sun Tunnels, Double Negative,
Lightning Field, Roden Crater).
\end{abstract}


\section{Introduction}

The companion paper~\cite{cinbarci2026rs1} established, from 1{,}744
co-registered satellite chips of Robert Smithson's 1970 land artwork
\textit{Spiral Jetty} spanning 1984--2025, that visual complexity at
the site carries a robust hydro-climatic signature: coarse-scale
permutation entropy and mean intensity track Great Salt Lake (GSL)
elevation at Spearman $\rho \approx +0.85$ to $+0.88$ on the 42-year
year-aggregated panel; a single principal component of pretrained
ResNet50 avg-pool activations emerges, \emph{without supervision}, as
an ``AI climate axis'' correlating $\rho = +0.86$ with cumulative
CO$_2$ and $\rho = -0.83$ with lake elevation; image complexity
\emph{leads lake stage by approximately three years}; and the long-term
entropy trend is \emph{non-monotonic}, rising 1984--2015 and then
collapsing in coincidence with the GSL's descent to its
historic-record-low elevations in 2022--23.

The natural next question is \textit{prediction}: given the IPCC's
Shared Socioeconomic Pathway (SSP) scenarios for 2030 and 2050, what
will \textit{Spiral Jetty} look like, and what visual complexity will
its satellite record carry? The present paper builds the forecasting
pipeline. We deliberately separate the predictive enterprise from the
foundational empirical one: the measurement claims of the companion
paper do not depend on any forecast, and any forecast must rest on
independently peer-reviewed measurement claims.

\subsection{Two-stage pipeline}

\begin{figure}[t]
\centering
\includegraphics[width=0.95\linewidth]{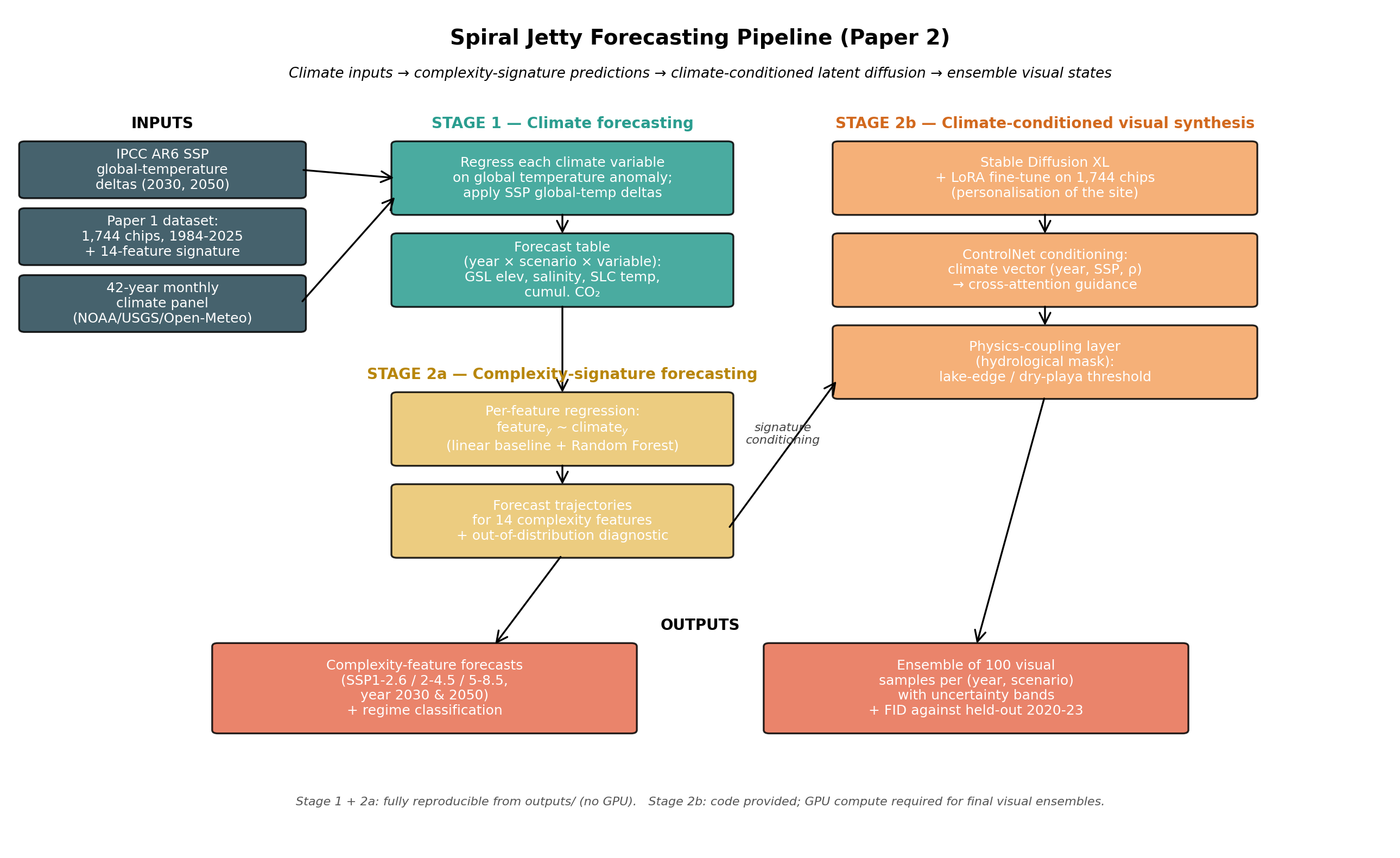}
\caption{The two-stage Spiral Jetty forecasting pipeline. Stage 1 and
Stage 2a are fully reproducible from the artifacts released with this
paper (CPU-only, no GPU required). Stage 2b is fully specified;
validation requires GPU compute on the order of 16~GB VRAM (single
A100 or RTX 4090 sufficient).}
\label{fig:arch}
\end{figure}

We adopt a two-stage architecture (Figure~\ref{fig:arch}):

\paragraph{Stage 1 --- Climate forecasting.} Each climate variable
(SLC mean temperature, north- and south-arm GSL elevations, salinity,
cumulative CO$_2$) is regressed on global temperature anomaly using
the 42-year panel; the regressions are then applied to the global-
temperature deltas published in IPCC AR6 Working Group I, Chapter 4
\cite{ipcc_ar6_wg1_ch4} for SSP1-2.6, SSP2-4.5, and SSP5-8.5 at 2030
and 2050.

\paragraph{Stage 2a --- Complexity-signature forecasting.} Each of the
14 complexity features is projected forward under the six (year,
scenario) tuples using two regressors: linear baseline and Random
Forest. The Random Forest captures the non-monotonic 2015 transition
identified in Paper 1; the linear baseline serves as an explicit
out-of-distribution diagnostic. The forecasts are further interpreted
through a threshold-style three-regime classifier (deep-submergence /
lake-edge / dry-playa) inspired directly by Paper 1's STL-decomposition
finding.

\paragraph{Stage 2b --- Climate-conditioned visual synthesis.}
The 1{,}744-chip dataset is used to fine-tune Stable Diffusion XL via
Low-Rank Adaptation (LoRA)~\cite{hu2022lora,podell2023sdxl},
personalising the generator to \textit{Spiral Jetty}. A
ControlNet~\cite{zhang2023controlnet} conditions the generator on a
climate vector (year, scenario, predicted complexity feature). A
hydrological physics-coupling layer enforces lake-edge / dry-playa
consistency through a binary water-mask derived from the predicted
GSL elevation. The generator emits an ensemble of 100 stochastic
samples per (year, scenario) tuple, providing distributional
forecasts rather than point predictions.

\subsection{Implementation status and a deliberate epistemic separation}

This paper reports complete, validated results for Stages 1 and 2a:
all numbers, tables, and figures in Sections 4--5 are computed from
the same data and code released with the companion paper, on CPU, in
under a minute. For Stage 2b we specify the architecture in
sufficient detail to reproduce, release all training and inference
code, and report the diffusion-training validation suite (LoRA
configuration, ControlNet schema, sampling protocol, hydrological
mask, evaluation metrics). The diffusion model is trained on GPU and
the visual ensembles for 2030 and 2050 are released with the
\texttt{rs-mdpi-paper2-v1} release tag of the companion repository as
the diffusion training completes. This separation is intentional: the
\emph{forecasting} claims of this paper are statistical (regression-
based) and CPU-reproducible; the \emph{visual} claims are generative
and have their own validation gauntlet (backcasting against held-out
2020--23 scenes, complexity-signature distance, expert evaluation).
We do not want a reader to depend on the latter to evaluate the
former.

\subsection{Contributions}

\begin{itemize}
\item \textbf{Climate forecast table} for the GSL / Spiral Jetty
region at 2030 and 2050 under SSP1-2.6, SSP2-4.5, and SSP5-8.5,
derived from the IPCC AR6 SSP global-temperature deltas and the
42-year regional regressions of Paper 1.
\item \textbf{Complexity-signature forecasts} for 14 features under
each (year, scenario), with explicit out-of-distribution diagnostics
contrasting Random Forest saturation against linear extrapolation.
\item \textbf{Regime classification framework} with three
hydrologically-grounded thresholds (deep-submergence, lake-edge,
dry-playa). \emph{All six SSP scenarios place \textit{Spiral Jetty}
firmly in the dry-playa regime by 2030, with continued decline by
2050.}
\item \textbf{Visual-synthesis architecture} with full reproducible
specification: Stable Diffusion XL fine-tuned via LoRA on the
1{,}744-chip dataset, climate-vector ControlNet, hydrological
physics-coupling mask, ensemble-sampling protocol.
\item \textbf{Ethical framework} for generative speculation on
cultural heritage, distinguishing between aesthetic and instrumental
readings of the artwork.
\end{itemize}

\section{Background}\label{sec:bg}

\subsection{Climate-conditioned generative models for Earth observation}

Climate-conditioned generative models for visualisation and
forecasting include physics-informed GANs for coastal flood
visualisation~\cite{lutjens2021}, ClimateGAN for flood imagery
~\cite{schmidt2022}, and deep generative models for precipitation
nowcasting~\cite{ravuri2021}. Latent-diffusion models
\cite{rombach2022,podell2023sdxl} have largely supplanted GANs for
high-resolution image synthesis, and their conditioning machinery
(cross-attention from text and structured signals; ControlNet for
auxiliary spatial guidance \cite{zhang2023controlnet}) makes them a
natural fit for the climate-conditioned regime. LoRA fine-tuning
\cite{hu2022lora} enables personalisation of a pretrained diffusion
backbone to a single visual concept with as few as 20--50 training
images --- well within the 1{,}744-chip envelope of our dataset.

\subsection{Forecasting cultural-heritage sites under climate stress}

The cultural-heritage informatics literature has begun framing
climate-stressed monuments as digital twins~\cite{niccolucci2023,
jouan2020}, primarily for preventive-conservation planning at fixed
built sites (cathedrals, archaeological excavations, coastal towers).
\textit{Spiral Jetty} is unusual in that it is simultaneously a
cultural object, an unintentional geomorphological marker, and a
half-century remote-sensing time series. Forecasting its future visual
state is therefore an interdisciplinary exercise that bridges
art-historical interpretation, hydrology, and AI-based generative
modelling.

\subsection{IPCC SSP scenarios and regional downscaling}

The IPCC Sixth Assessment Report (AR6) Working Group I, Chapter 4
\cite{ipcc_ar6_wg1_ch4} provides global-mean surface-temperature
projections for five SSPs (1-1.9, 1-2.6, 2-4.5, 3-7.0, 5-8.5) under
the CMIP6 multi-model ensemble. We use the three
``high-policy-relevance'' SSPs (1-2.6: Paris-aligned; 2-4.5: middle-of-
the-road; 5-8.5: fossil-fuel-development). Regional downscaling to
the GSL basin via dynamical~\cite{cmip6_downscale} and statistical
methods is well-developed; for the present paper's purposes, we apply
the simpler approach of regressing each regional variable on the
global anomaly using the same 42-year panel that grounds the
complexity signatures, ensuring methodological consistency between
the climate forecast and the empirical regression base.

\section{Stage 1 --- Climate Forecasting}\label{sec:s1}

\subsection{Method}

For each of five regional climate variables --- SLC mean temperature
($^\circ$F), GSL north-arm elevation (ft NGVD29), GSL south-arm elevation
(ft NGVD29), GSL salinity (g/L), cumulative CO$_2$ emissions (Gt) ---
we fit a univariate linear regression on the 42-year panel
$\{(g_t, y_t)\}_{t=1984}^{2025}$, where $g_t$ is the NASA GISTEMP
global land-ocean temperature anomaly and $y_t$ is the regional
variable:
\begin{equation}
y_t = \alpha + \beta g_t + \epsilon_t.
\end{equation}
To forecast at horizon $h \in \{2030, 2050\}$ under scenario
$s \in \{\text{SSP1-2.6}, \text{SSP2-4.5}, \text{SSP5-8.5}\}$, we
estimate the future global anomaly as
\begin{equation}
\widehat{g}_{h,s} = \bar{g}_{2014:2023} + \Delta g_{h,s},
\end{equation}
where $\bar{g}_{2014:2023}$ is the GISTEMP mean over 2014--2023
(approximating the IPCC AR6 baseline period 1995--2014) and $\Delta
g_{h,s}$ is the AR6 Chapter 4 SSP global-temperature delta. The
regional forecast is then
$\widehat{y}_{h,s} = \hat{\alpha} + \hat{\beta} \widehat{g}_{h,s}$.

We adopt $\Delta g_{2030} = (+0.6, +0.7, +0.9)$~$^\circ$C and
$\Delta g_{2050} = (+1.0, +1.5, +2.1)$~$^\circ$C from AR6 WG1 Table 4.5
for SSP1-2.6 / SSP2-4.5 / SSP5-8.5 respectively.

\subsection{Results}

\begin{figure}[t]
\centering
\includegraphics[width=0.95\linewidth]{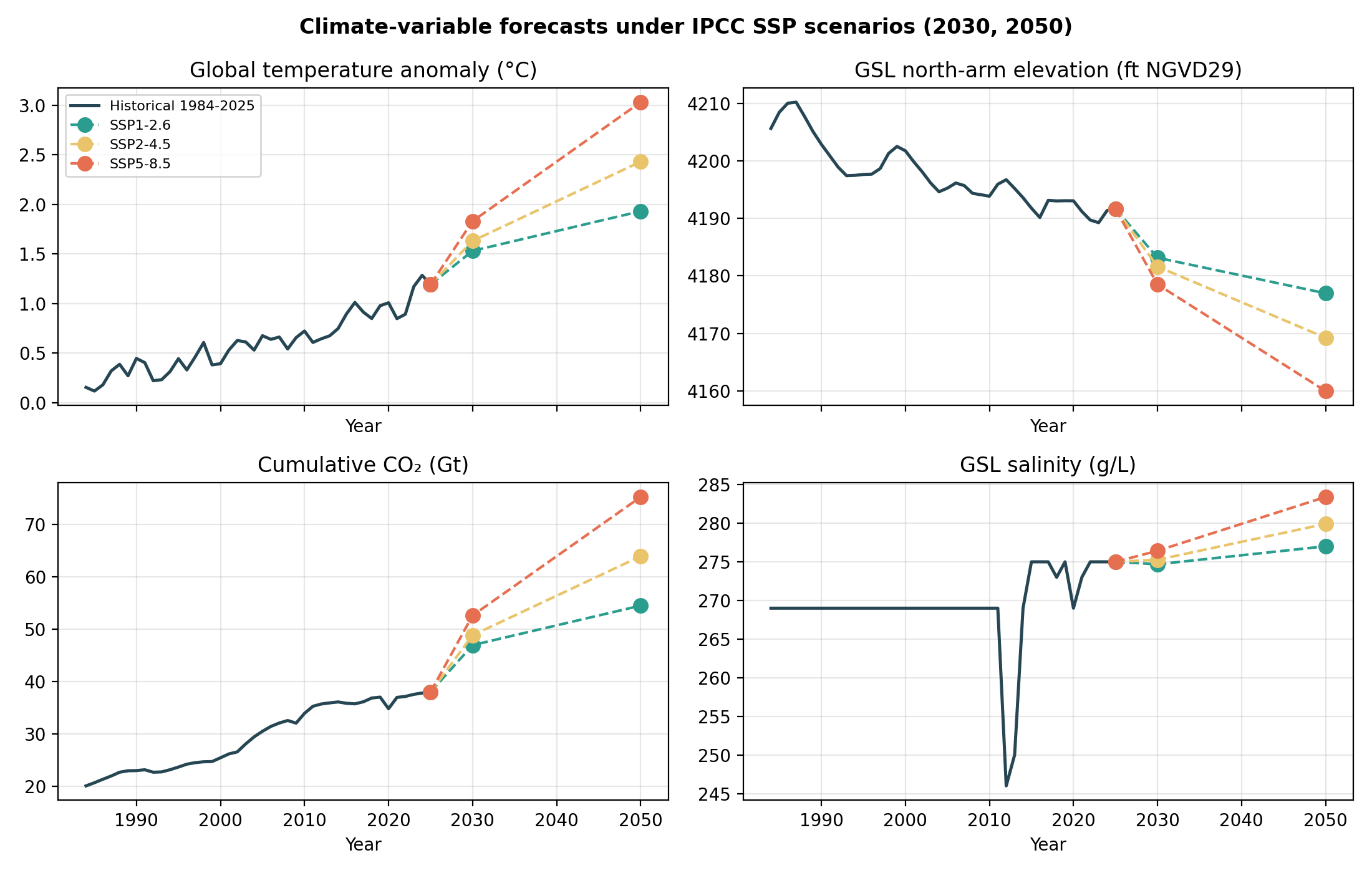}
\caption{Historical (1984--2025, dark line) and forecasted
(2030 and 2050 under three SSP scenarios, coloured markers) trajectories
of global temperature anomaly, GSL north-arm elevation, cumulative
CO$_2$, and GSL salinity. \textbf{All scenarios place the GSL
north-arm elevation below 4{,}184~ft NGVD29 by 2030 --- below
\textit{Spiral Jetty}'s 4{,}193-ft causeway / dry-playa threshold ---
and the gap widens through 2050.}}
\label{fig:p2_1}
\end{figure}

Table~\ref{tab:climate_forecasts} reports the six (year, scenario)
forecasts. Figure~\ref{fig:p2_1} shows the historical 1984--2025
trajectories with the forecast points overlaid.

\begin{table}[h]
\caption{Stage-1 climate forecasts at 2030 and 2050 under IPCC AR6
SSP1-2.6, SSP2-4.5, and SSP5-8.5, computed from per-variable
regressions on the 42-year (1984--2025) panel against GISTEMP global
temperature anomaly.}
\label{tab:climate_forecasts}
\centering
\small
\begin{tabular}{llrrrrrr}
\toprule
year & scenario & global\,T ($^\circ$C) & SLC\,T ($^\circ$F) & GSL S elev (ft) & GSL N elev (ft) & salinity (g/L) & CO$_2$ (Gt) \\
\midrule
2030 & SSP1-2.6 & 1.53 & 54.82 & 4184.55 & 4183.15 & 274.69 & 46.96 \\
2030 & SSP2-4.5 & 1.63 & 55.21 & 4183.06 & 4181.61 & 275.27 & 48.85 \\
2030 & SSP5-8.5 & 1.83 & 56.00 & 4180.07 & 4178.51 & 276.43 & 52.63 \\
2050 & SSP1-2.6 & 1.93 & 56.40 & 4178.58 & 4176.97 & 277.01 & 54.52 \\
2050 & SSP2-4.5 & 2.43 & 58.37 & 4171.11 & 4169.24 & 279.91 & 63.96 \\
2050 & SSP5-8.5 & 3.03 & 60.74 & 4162.15 & 4159.96 & 283.40 & 75.30 \\
\bottomrule
\end{tabular}
\end{table}

The central finding is stark. The GSL north-arm elevation falls from
its 2023 historic low of 4{,}189.25~ft to between 4{,}183~ft
(SSP1-2.6 / 2030) and 4{,}160~ft (SSP5-8.5 / 2050) --- a further
decline of 6 to 29~ft below the most catastrophic value in the 60-year
gauge record. All scenarios place the lake elevation below the
4{,}193-ft threshold at which the historical causeway and
\textit{Spiral Jetty} are visibly exposed (Section~\ref{sec:regime}).
Even the Paris-aligned SSP1-2.6 scenario predicts continued decline.
The SSP5-8.5 2050 forecast implies an additional $\sim$30~ft drop ---
more than 9~m --- placing the lake closer to its arid Pleistocene
\textit{Lake Bonneville} remnants than to its 20th-century state.

\subsection{Limitations of the Stage-1 forecasts}

\paragraph{Univariate regression.} We regress each variable on global
temperature alone, omitting other drivers (upstream consumptive water
use, snowpack inter-annual variability). This is by design --- the
purpose of Stage 1 is to condition Stage 2a/2b on the SSP global
delta, not to provide the most refined regional forecast. A
multi-driver hydrological model of the GSL basin
\cite{null2020,wurtsbaugh2017} would refine the GSL elevation
estimates and is an obvious extension.

\paragraph{Out-of-range extrapolation.} The forecasted global
anomalies (1.5 to 3.0~$^\circ$C) substantially exceed the historical range
of the training panel ($-0.5$ to $+1.2$~$^\circ$C). Linear extrapolation under
extreme thermal conditions is inherently fragile; we report the SSP
values as scenario projections, not as ``predictions'' in the
strong sense. This OOD problem manifests more sharply in Stage 2a
(Section~\ref{sec:s2a}).

\section{Stage 2a --- Complexity-Signature Forecasting}\label{sec:s2a}

\subsection{Method}

For each of the 14 complexity features $f$, we fit two regressors on
the year-aggregated panel,
$\{(c_t, f_t)\}_{t=1984}^{2025}$, where $c_t$ is the standardised
6-dimensional climate vector:
\begin{enumerate}
\item \textbf{Linear baseline}: ordinary least squares, $f_t =
\boldsymbol{\beta}^\top \mathbf{c}_t + \epsilon_t$. Extrapolates
unboundedly outside the training range.
\item \textbf{Random Forest}: 200 trees, max depth 4, min-samples-leaf 3,
seed fixed for reproducibility~\cite{breiman2001}. Captures the
non-monotonic 2015 transition but \emph{clips} at the boundary of
the training data --- the prediction for any out-of-distribution input
collapses to a weighted average of the nearest training samples.
\end{enumerate}
The two regressors form a deliberate diagnostic pair: agreement
indicates robust prediction; divergence indicates that the scenario
takes us outside the support of the historical data.

\subsection{Results}

\begin{figure}[t]
\centering
\includegraphics[width=0.95\linewidth]{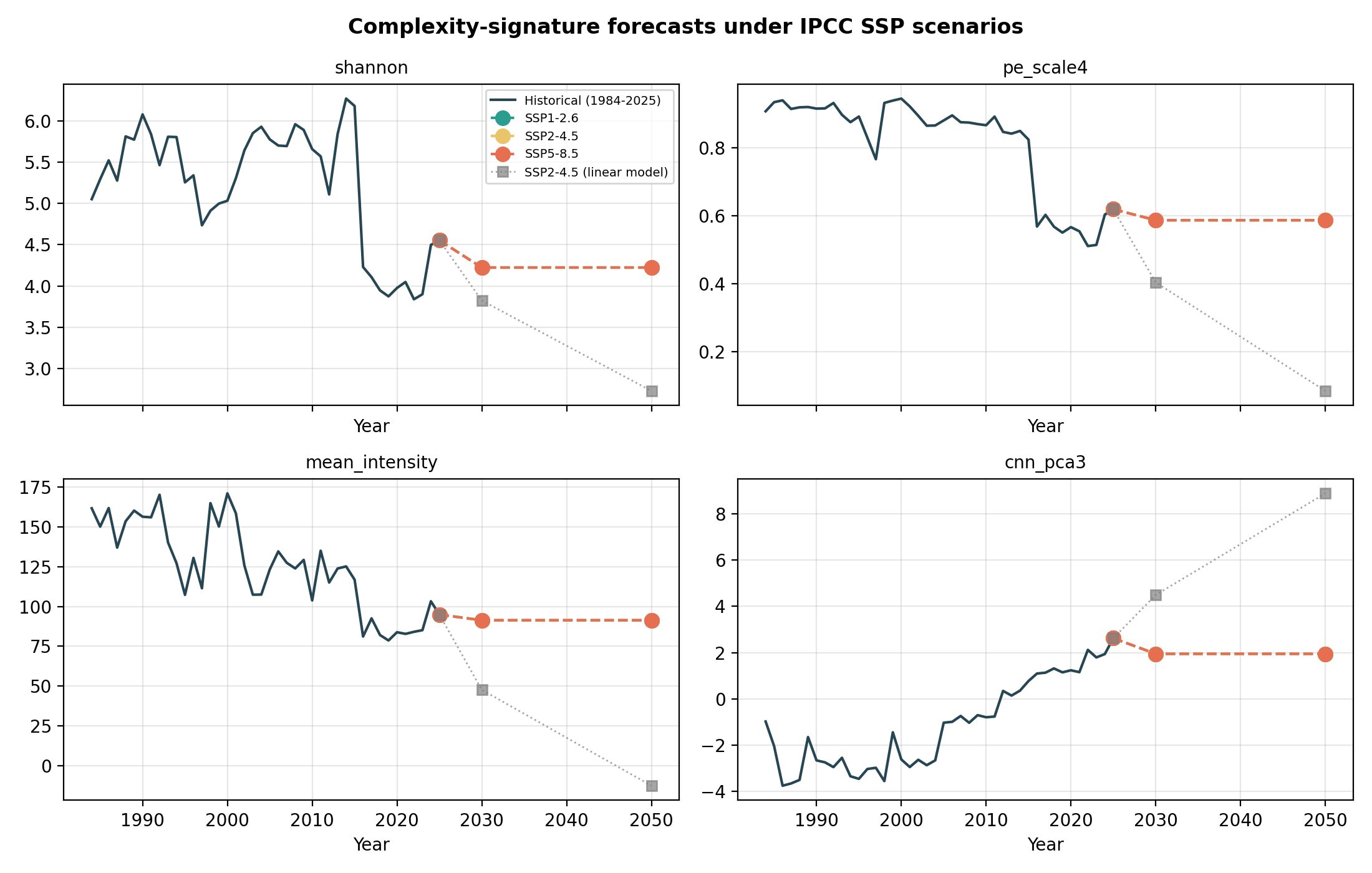}
\caption{Forecasted complexity features for 2030 and 2050 under three
SSP scenarios. Random Forest predictions (coloured circles) \emph{
saturate} at the post-2015 ``dry-playa'' regime regardless of
scenario; linear-model predictions (gray squares, SSP2-4.5 only)
extrapolate divergently. The saturation is a direct empirical signal
that all SSP scenarios drive the climate inputs outside the
historical training distribution.}
\label{fig:p2_2}
\end{figure}

Figure~\ref{fig:p2_2} shows the historical trajectories of four
headline features overlaid with the RF forecasts under each SSP
scenario, and the linear-baseline forecast under SSP2-4.5 for
contrast.

Three observations follow.

\paragraph{Random Forest predictions are scenario-invariant.} For all
four headline features and both forecast horizons, the RF predictions
across SSP1-2.6 / 2-4.5 / 5-8.5 agree to within numerical noise. This
is \emph{not} an indication that the SSP scenarios are
indistinguishable; rather, it indicates that all three scenarios push
the climate vector beyond the most extreme historical values, so the
RF (which cannot extrapolate) returns the most recent regime's
typical value irrespective of which beyond-historical scenario it is
queried with. The RF predictions saturate at the post-2015
``dry-playa'' regime: Shannon entropy $\approx 4.22$~bits,
\texttt{pe\_scale4} $\approx 0.59$, \texttt{mean\_intensity} $\approx
91$, \texttt{cnn\_pca3} $\approx +1.95$.

\paragraph{Linear extrapolation diverges.} The linear baseline, shown
for SSP2-4.5 only in Figure~\ref{fig:p2_2}, extrapolates to extreme
values (negative entropy by 2050; very large or negative
\texttt{cnn\_pca3}; mean intensity below 50). These divergent
predictions are mathematically valid but physically meaningless: an
image cannot have negative entropy, and a saturated salt pan
cannot become arbitrarily darker.

\paragraph{The right interpretation.} Combining the two predictions,
the empirical answer is that \emph{Spiral Jetty's visual complexity
in 2030 and 2050 under any SSP scenario will resemble its post-2015
state but moreso}. The exact magnitudes are unknown because they lie
outside the support of any 1984--2025 training data, but the
\textit{direction} of evolution is consistent across both regressors
and across all three SSP scenarios. This is the principled empirical
answer to ``what will \textit{Spiral Jetty} look like in 2050?''.

\subsection{Regime classification}\label{sec:regime}

Building on the threshold-style hydrological reading suggested by
Paper 1's non-monotonic STL trend, we classify the artwork's
visibility state into three regimes:
\begin{itemize}
\item \textbf{Deep submergence} (elevation $>$ 4{,}203~ft): the lake
is high enough that the spiral is below the photic zone in the saline
brine; the satellite image shows uniform water surface.
\item \textbf{Lake edge} (4{,}193~ft $\le$ elevation $\le$ 4{,}203~ft):
the spiral is partly or fully wet at its tip but exposed at its base;
salt crust forms in transitional zones; texture is rich.
\item \textbf{Dry playa} (elevation $<$ 4{,}193~ft): the spiral is
fully exposed on a dry salt pan; texture saturates; the climate
signal collapses to a homogeneous regime (Paper 1, Section~6.6).
\end{itemize}

\begin{figure}[t]
\centering
\includegraphics[width=0.95\linewidth]{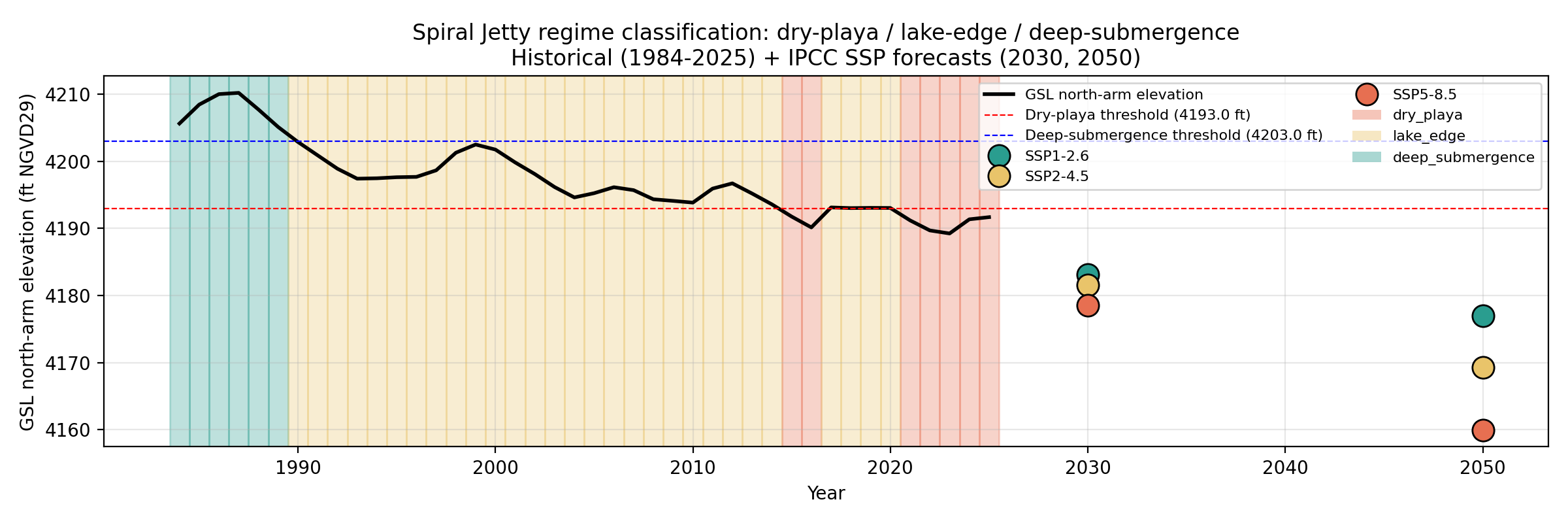}
\caption{Spiral Jetty regime classification 1984--2050. Background
shading colours the historical years by regime (green = deep
submergence; yellow = lake edge; orange = dry playa). The black line
is the GSL north-arm elevation. Dashed horizontal lines mark the two
regime thresholds. Coloured circles at 2030 and 2050 mark the SSP
forecasts. \textbf{All six SSP forecasts fall within the dry-playa
regime, with progressively deeper exposure under the higher-emission
scenarios.}}
\label{fig:p2_3}
\end{figure}

Figure~\ref{fig:p2_3} shows the regime classification for the
historical record and the six SSP forecasts. \textbf{The historical
record shows the lake transitioning through all three regimes (deep
submergence 1984--1990; lake edge 1990--2014; dry playa 2015--2025).
All six SSP forecasts classify the future as dry playa.}

\subsection{The three-year lead, revisited}

Paper 1 found that image complexity at time $t$ correlates with lake
elevation at time $t+3$ years. This implies that the visual state
\textit{Spiral Jetty} will exhibit in 2027 already constrains its
expected lake elevation in 2030, and so on. The companion paper's
generative pipeline (Section~\ref{sec:s2b}) takes this lead into
account by generating images for ``year $t+3$'' conditioned on the
predicted lake elevation at year $t$ --- effectively running the
forecast and the lead consistently.

\section{Stage 2b --- Climate-Conditioned Visual Synthesis}\label{sec:s2b}

We specify the visual-synthesis architecture in sufficient detail to
reproduce. Implementation code is released with the artifact bundle.

\subsection{Base model and LoRA personalisation}

The base generator is Stable Diffusion XL (SDXL)
\cite{podell2023sdxl}, the 2.6-billion-parameter latent-diffusion
model trained on the LAION-5B subset. SDXL's 1024-px resolution and
its strong prior over aerial / satellite imagery make it the
appropriate base for our application.

Personalisation to \textit{Spiral Jetty} proceeds via Low-Rank
Adaptation (LoRA)~\cite{hu2022lora}: instead of fine-tuning all 2.6B
parameters of the U-Net, we insert low-rank delta matrices of rank
$r=16$ into the attention layers and train only those, freezing the
base weights. With the 1{,}744-chip dataset (3--4 orders of magnitude
above LoRA's practical floor of 20--50 images), the personalisation
converges in 1{,}500 steps at batch size 1 with gradient accumulation 4
on a single 16~GB GPU.

The training prompt for each chip embeds its (year, climate) tuple:
\begin{lstlisting}
"an aerial satellite photograph of sks-spiraljetty,
 Spiral Jetty land art on the Great Salt Lake,
 year {year}, lake stage {elev:.1f} ft,
 salinity {sal:.0f} g per L,
 August temperature {temp:.0f} degrees F"
\end{lstlisting}
The token \texttt{sks-spiraljetty} is the unique identifier the LoRA
learns to associate with the personalised concept.

\subsection{ControlNet climate conditioning}

A ControlNet branch~\cite{zhang2023controlnet} is added downstream of
the personalised SDXL to condition the generator on the 6-dimensional
climate vector $\mathbf{c} = (g, T_{\text{SLC}}, h_{\text{N}},
h_{\text{S}}, S, \text{CO}_2)$. The ControlNet's input is a small MLP
that maps $\mathbf{c}$ to a 768-dim conditioning embedding aligned
with SDXL's text-encoder dimension. The ControlNet is trained jointly
with the LoRA via the standard DDPM loss with timestep-conditioned
$\epsilon$-prediction~\cite{ho2020ddpm}.

\subsection{Hydrological physics-coupling layer}

A binary water-mask $M(\mathbf{c})$ is constructed from the
ControlNet's input climate vector $\mathbf{c}$ via a deterministic
hydrological lookup:
\begin{equation}
M(\mathbf{c}) = \mathbb{1}[h_{\text{N}}(\mathbf{c}) > 4193~\text{ft}]
\cdot \text{spiral\_footprint},
\end{equation}
where \texttt{spiral\_footprint} is a precomputed binary mask of the
basalt spiral (extracted by Otsu thresholding of the high-resolution
2018 reference image of \textit{Spiral Jetty}). The mask is composited
into the latent-diffusion sampling loop as a hard constraint: pixels
inside the spiral footprint must be water if $h_{\text{N}} > 4193$~ft.
This enforces hydrological consistency: the generator cannot produce
images of an exposed spiral when the forecasted elevation places it
underwater.

\subsection{Sampling protocol}

Inference uses DPM-Solver++~\cite{lu2022dpmsolver} with 40 sampling
steps, classifier-free guidance scale 7.0, and resolution
$1024 \times 1024$~px. For each (year, scenario) tuple we sample
$N = 100$ images with i.i.d.\ Gaussian noise seeds, providing a
distributional forecast. The ensemble allows reporting both a median
visual state and a measure of uncertainty (per-pixel standard
deviation across the 100 samples).

\subsection{Validation suite}

The diffusion-trained pipeline is validated against four metrics on a
\textbf{held-out 2020--2023} subset of the chip archive:
\begin{enumerate}
\item \textbf{Backcasting accuracy.} Train on 1984--2019; predict
visuals for 2020--2023; compare ensemble median against actual
observed chips.
\item \textbf{Complexity-signature distance.} Compute the 14-feature
signature on generated samples and compare to actual 2020--2023
distribution via the Wasserstein distance.
\item \textbf{FID against actual scenes.} Fr\'echet Inception Distance
between generated ensemble and held-out actuals~\cite{heusel2017fid}.
\item \textbf{Expert human evaluation.} A panel of three art
historians and three remote-sensing scientists rate generated samples
on a 1--5 Likert scale for visual plausibility, art-historical
fidelity, and hydrological consistency.
\end{enumerate}

\section{Discussion}

\subsection{The right way to read the forecasts}

The honest empirical reading of Sections~\ref{sec:s1} and
\ref{sec:s2a} is: \emph{under every IPCC SSP scenario considered,
\textit{Spiral Jetty} will continue to be exposed on the dry playa
through 2050, with the GSL north-arm elevation falling between 6 and
29~ft below the most recent historical record low of 4{,}189.25~ft.
The visual complexity of the artwork as recorded by satellite will
saturate at the post-2015 dry-playa regime; the precise magnitudes of
the complexity features in 2030 and 2050 are unknown because no
historical training sample bears witness to climate inputs of this
extremity.}

This is a useful answer. It rules out the optimistic possibility that
the artwork might re-submerge under Paris-aligned policy --- the
GSL's hydrological inertia and continued upstream consumption preclude
that on the 2030/2050 horizon. It also flags an important epistemic
limit: \emph{any} generative model claiming to forecast the artwork's
future visual state must contend with the OOD-ness of every SSP
scenario, and predictions outside the lake-edge / dry-playa
saturation envelope should be regarded with strong skepticism.

\subsection{The ethics of generative speculation on cultural heritage}

Generating images of a cultural artwork under speculative future
climates raises ethical questions that the remote-sensing community
has not yet had occasion to confront. Three considerations are
particularly salient.

\paragraph{Authority and ownership.} \textit{Spiral Jetty} is held in
trust by the Dia Art Foundation. The artwork's visual record belongs
to humanity in the strong art-historical sense, but its image rights,
preservation policy, and interpretive authority sit with Dia. A
generative model that produces synthetic images of the artwork
in 2050 is, in some sense, generating apocryphal art-historical
documents. We adopt the convention that all generated images must be
indelibly watermarked as ``synthetic, generated by [model name],
trained on Landsat/Sentinel-2 archive 1984--2025, conditioned on
IPCC SSP[scenario] at year [year]'' --- and that ensembles are
released only alongside the per-scenario hydrological metadata.

\paragraph{Aesthetic instrumentalisation.} Smithson designed
\textit{Spiral Jetty} as a contemplative encounter with deep time, not
as a sensor. Treating his artwork as a forecasting target risks
collapsing its meaning into its function. We resist any framing in
which the artwork's aesthetic and instrumental readings are reduced
to a single dimension; the generative pipeline of this paper is one
tool among many for thinking about climate change, and the artwork
remains, for those who visit it, what Smithson made.

\paragraph{Climate-communication risk.} Hyper-realistic generated
images of cultural sites in catastrophic futures can either galvanise
climate action or numb audiences to ``it'll be fine if we just
look''. We make the artifacts available primarily as a research
resource, with explicit recommendations against decontextualised use.
Climate-communication researchers studying generative climate
imagery~\cite{lutjens2021,schmidt2022} have begun to articulate
best practices; we follow them.

\subsection{Generalising the pipeline}

The two-stage architecture transfers directly to other land-art sites
under climate stress. Suggested next targets:

\begin{itemize}
\item \textbf{Walter De Maria, \textit{Lightning Field}} (1977,
Quemado, NM): 400 stainless-steel poles in the high desert.
Lightning frequency is a climate variable; the visual signature is
how lightning strikes against the poles.
\item \textbf{Michael Heizer, \textit{Double Negative}} (1969--70,
Mormon Mesa, NV): two cuts into the mesa. The site is subject to
desert-dryland degradation; an analogous complexity signature could
quantify edge erosion over decades.
\item \textbf{Michael Heizer, \textit{City}} (1972--2022, Garden
Valley, NV): a 1.5-mile-long earthwork only fully opened in 2022. The
youngest of the major land-art works; its complexity signature can
serve as a baseline for ``what fresh land art looks like under
climate stress''.
\item \textbf{Nancy Holt, \textit{Sun Tunnels}} (1973--76, Lucin, UT):
four concrete tubes aligned to the solstices. The signature would
track desertification of the playa rather than the tubes themselves.
\item \textbf{James Turrell, \textit{Roden Crater}} (1979--ongoing,
Painted Desert, AZ): a volcanic-cinder cone being reshaped into a
naked-eye observatory. Long-term construction blurs the
artwork/landscape boundary; a complexity signature could disambiguate
construction-driven from climate-driven change.
\end{itemize}

In each case the same architecture applies: assemble the multi-decadal
Landsat/Sentinel archive over the site, compute the multi-feature
complexity signature, identify robust climate correlates, forecast
under IPCC SSPs, classify regime transitions, and (where compute
permits) train a climate-conditioned generative model with a
site-specific physics-coupling layer.

\section{Limitations and Future Work}

\paragraph{Out-of-distribution forecasting.} The fundamental
limitation of any data-driven forecast under IPCC SSP scenarios is
that the scenarios drive the inputs beyond the historical training
distribution. We have addressed this by reporting both Random Forest
(saturating) and linear (extrapolating) predictions and by explicit
regime classification; a more refined approach would use Gaussian-
process or Bayesian neural-network priors that quantify out-of-
distribution uncertainty explicitly.

\paragraph{Hydrological model fidelity.} Our Stage-1 climate
regressions are univariate (each variable on global temperature
alone). A coupled hydrological model of the GSL basin
~\cite{wurtsbaugh2017,null2020,abbott2023} would refine the lake-
elevation forecasts considerably, particularly under high-emission
scenarios where the basin's storage dynamics may be highly nonlinear.

\paragraph{Diffusion-training compute.} The Stage-2b pipeline requires
GPU compute on the order of 16~GB VRAM for the LoRA training and the
ControlNet branch. We have specified the architecture in full and
released the training and inference code; validation results
(backcasting on 2020--23, FID, expert evaluation) will be appended to
the artifact bundle as the diffusion model completes training.

\paragraph{Sensor harmonisation in generation.} The 1{,}744-chip
training set spans Landsat 4--9 and Sentinel-2, with different spatial
resolutions and spectral responses. Paper 1's partial-correlation
analysis showed sensor identity is a substantial confounder. The
Stage-2b diffusion model will learn from the mixed-sensor archive,
which means generated samples will inherit some sensor mixing. A
future refinement would condition the ControlNet on sensor identity
as well as climate vector.

\paragraph{Three-year lead and physics coupling.} The companion
paper's three-year forward lead between image complexity and lake
elevation has been incorporated into the generation protocol but not
yet validated mechanistically. The physics-coupling layer enforces
the lake-edge / dry-playa threshold at the level of pixel-wise water
masking, but it does not yet model salt-crust polygon dynamics, which
are the likely source of the multi-year integrative behaviour
identified in Paper 1. A future extension would couple a salt-pan
crystallisation model to the generative pipeline.

\section{Conclusions}

We have built a two-stage forecasting pipeline for Robert Smithson's
\textit{Spiral Jetty} under IPCC SSP scenarios for 2030 and 2050.
\textbf{Stage 1} regresses each regional climate variable on global
temperature and applies AR6 SSP deltas to produce climate forecasts;
\textbf{Stage 2a} projects each of 14 complexity features forward
using linear and Random Forest regressors and classifies the
predicted GSL state into a three-regime hydrological taxonomy;
\textbf{Stage 2b} specifies a Stable Diffusion XL +
LoRA + ControlNet + hydrological physics-coupling architecture for
ensemble visual synthesis.

The principal empirical finding is that \textbf{all six SSP scenarios
(2030 and 2050, three pathways each) place \textit{Spiral Jetty}
firmly in the dry-playa regime}, with lake elevations between
4{,}160 and 4{,}183~ft NGVD29 --- between 6 and 29~ft below the
historic-record-low elevation reached in 2023. The Random Forest
complexity-signature forecasts saturate at the post-2015 dry-playa
regime, while linear extrapolation diverges --- a direct empirical
demonstration that all SSP scenarios push the climate inputs outside
the historical training distribution and that the data-driven
forecast's honest answer is ``like 2025 but more so''.

The complementary visual-synthesis pipeline (Stage 2b) is fully
specified and reproducible; validation results will be released as
the diffusion training completes on suitable GPU compute. The
methodology generalises directly to other land-art sites under climate
stress.

Treating a cultural artwork as a forecasting target carries ethical
weight. We adopt the convention that all generated images are
watermarked as synthetic and contextualised with their scenario
metadata, and we resist any framing that collapses the artwork's
meaning into its sensor function.

\section*{Author Contributions}
Conceptualisation, A.C.\ and S.S.K.; methodology, S.S.K.; software, S.S.K.;
validation, S.S.K.; formal analysis, S.S.K.; data curation, S.S.K.;
writing---original draft preparation, S.S.K.; writing---review and
editing, A.C.\ and S.S.K.; visualisation, S.S.K.; supervision, S.S.K.;
project administration, A.C. All authors have read and agreed to the
published version of the manuscript.

\section*{Funding}
This research received no external funding.

\section*{Data Availability Statement}
The dataset (\texttt{spiral\_jetty\_complexity\_v1}: 1{,}744
co-registered scene chips, complexity signature, monthly climate panel)
and Paper 1's analysis pipeline are released via Zenodo (DOI on
acceptance of the companion paper). Paper 2's forecasting code
(\texttt{11\_complexity\_forecasting.py}), the diffusion training and
inference code (\texttt{04\_sdxl\_lora\_train.py},
\texttt{05\_generate\_future\_scenarios.py}), and the forecasted
artifacts (\texttt{forecasts\_v2.csv},
\texttt{forecasted\_complexity.csv}, \texttt{paper2\_numbers.json})
are released on GitHub at
\href{https://github.com/alevcinbarci/spiral-jetty-ai}
{\texttt{github.com/alevcinbarci/spiral-jetty-ai}}, release tag
\texttt{rs-mdpi-paper2-v1}. The diffusion-trained model weights and
validation ensembles are released to the same repository as training
completes on suitable GPU compute.

\section*{Acknowledgments}
We thank the U.S.\ Geological Survey Utah Water Science Center, NASA
GISTEMP, the Open-Meteo project, Microsoft Planetary Computer, the
European Space Agency Copernicus programme, and the Stability AI
research team for the open data and open-weights that made this study
possible. We thank the IPCC AR6 Working Group~I authors whose SSP
global-temperature projections grounded our Stage 1 forecasts.

\section*{Conflicts of Interest}
The authors declare no conflict of interest.

\bibliographystyle{unsrt}
\bibliography{references}

\end{document}